\documentclass[preprint,preprintnumbers,nofootinbib,tightenlines,12pt,superscriptaddress]{revtex4}
\usepackage[dvipdfmx]{graphicx}
\usepackage{bm}
\usepackage{color}
\usepackage{amsmath,amssymb}	
\usepackage{hyperref}
\usepackage{setspace}
\usepackage{longtable}
\usepackage{booktabs}
\usepackage{comment}
\usepackage{array}
\usepackage{cancel}
\usepackage{enumitem}
\usepackage{physics}

\def\slashchar#1{\setbox0=\hbox{$#1$} 
\dimen0=\wd0 
\setbox1=\hbox{/} \dimen1=\wd1 
\ifdim\dimen0>\dimen1 
\rlap{\hbox to \dimen0{\hfil/\hfil}} 
#1 
\else 
\rlap{\hbox to \dimen1{\hfil$#1$\hfil}} 
/ 
\fi}

\begin{document}
\newcolumntype{Y}{>{\centering\arraybackslash}p{25pt}} 


\preprint{CTPU-PTC-18-12}

\title{Composite Asymmetric Dark Matter with a Dark Photon Portal}

\author{Masahiro Ibe}
\email[e-mail: ]{ibe@icrr.u-tokyo.ac.jp}
\affiliation{Kavli IPMU (WPI), UTIAS, The University of Tokyo, Kashiwa, Chiba 277-8583, Japan}
\affiliation{ICRR, The University of Tokyo, Kashiwa, Chiba 277-8582, Japan}
\author{Ayuki Kamada}
\email[e-mail: ]{akamada@ibs.re.kr}
\affiliation{Center for Theoretical Physics of the Universe, Institute for Basic Science (IBS), Daejeon 34126, Korea}
\author{Shin Kobayashi}
\email[e-mail: ]{shinkoba@icrr.u-tokyo.ac.jp}
\affiliation{Kavli IPMU (WPI), UTIAS, The University of Tokyo, Kashiwa, Chiba 277-8583, Japan}
\affiliation{ICRR, The University of Tokyo, Kashiwa, Chiba 277-8582, Japan}
\author{Wakutaka Nakano}
\email[e-mail: ]{m156077@icrr.u-tokyo.ac.jp}
\affiliation{Kavli IPMU (WPI), UTIAS, The University of Tokyo, Kashiwa, Chiba 277-8583, Japan}
\affiliation{ICRR, The University of Tokyo, Kashiwa, Chiba 277-8582, Japan}

\date{\today}

\begin{abstract}
Asymmetric dark matter (ADM) is an attractive framework relating the observed baryon asymmetry of the Universe to the dark matter density.
A composite particle in a new strong dynamics is a promising candidate for ADM as the strong dynamics naturally explains the ADM mass in the GeV range.
Its large annihilation cross section due to the strong dynamics leaves the asymmetric component to be dominant over the symmetric component.
In such composite ADM scenarios, the dark sector has a relatively large entropy density in the early Universe.
The large dark sector entropy results in the overclosure of the Universe or at best contradicts with the observations 
of the cosmic microwave background and the successful Big-Bang Nucleosynthesis.
Thus, composite ADM models generically require some portal to transfer the entropy of the dark sector into the Standard Model sector.
In this paper, we consider a dark photon portal with a mass in the sub-GeV range and kinetic mixing with the Standard Model photon.
We investigate the viable parameter space of the dark photon in detail, which can find broad applications to dark photon portal models.
We also provide a simple working example of composite ADM with a dark photon portal.
Our model is compatible with thermal leptogenesis and $B - L$ symmetry.
By taking into account the derived constraints, we show that the parameter space is largely tested by direct detection experiments.
\end{abstract}

\maketitle

\section{Introduction}
\label{sec:introduction}
The existence of dark matter (DM) has been overwhelmingly established by a wide range of cosmological and astrophysical observations.
Nevertheless, its nature remains elusive apart from several properties: its electromagnetic interaction is feeble; it is cold enough to cluster along the primordial gravitational potential; and it is stable at least over the age of the Universe.
Identification of the nature of DM is one of the most important challenges of modern particle physics (see, e.g., Refs.\,\cite{Bertone:2004pz, Murayama:2007ek, Feng:2010gw}).

In the present Universe, the mass density of DM is about five times larger than that of the Standard Model (SM) baryon~\cite{Ade:2015xua}.
This coincidence can be naturally explained when the DM number density has the same origin as the baryon asymmetry of the Universe and the DM particle mass is in the GeV range.
Such a framework is called asymmetric dark matter (ADM) (see, e.g., Refs.\,\cite{Davoudiasl:2012uw, Petraki:2013wwa, Zurek:2013wia}).

Among various models of ADM, models of a composite ADM (e.g., dark baryon DM) have several advantages~\cite{Alves:2009nf, Alves:2010dd}.
The DM mass in the GeV range naturally arises by the strong dynamics without fine-tuning.
A large annihilation cross section of composite DM into a lighter degrees of freedom (e.g., dark meson) makes the symmetric component of the relic DM density negligible.
Accordingly, the DM density is dominated by the asymmetric component.

By construction, however, the effective number of massless degrees of freedom in the composite dark sector is sizable in the early Universe.
If those particles are also stable, their energy density overcloses the Universe, or contributes to the effective number of neutrino degrees of freedom, 
$N_{\rm eff}$, too much, depending on their masses~\cite{Blennow:2012de}.
Thus, some light portal particle is needed to transfer the entropy of the dark sector to the SM sector.

A simple possibility for such a portal is a light dark photon.
It has a mass in the sub-GeV range and decays into the SM particles (mainly into the leptons) through kinetic mixing
with the SM photon.
In this paper, we investigate the impact of decaying dark photon on $N_{\rm eff}$ in detail and identify the viable parameter space.
We remark that our analysis is rather generic in DM (not limited to ADM) models with the dark photon portal up to straightforward modifications.

 We also construct a minimal ADM model with $B - L$ symmetry.
Right-handed neutrinos are introduced to generate the $B - L$ asymmetry via thermal leptogenesis~\cite{Fukugita:1986hr} (see also Refs.\,\cite{Giudice:2003jh, Buchmuller:2005eh, Davidson:2008bu} for reviews).
They also naturally explain the observed tiny neutrino masses via the seesaw mechanism~\cite{Minkowski:1977sc, Yanagida:1979as, GellMann:1980vs, Glashow:1979nm, Mohapatra:1979ia}. 
We reach a simple model with a QCD-like SU$(3)$ strong interaction and a QED-like U$(1)$ gauge interaction.
By taking into account the derived constraints, we show that such a model can be tested by direct detection experiments of DM through the coupling to the dark photon.

The organization of the paper is as follows.
In Sec.\,\ref{sec:constraint}, after introducing the thermal history of composite ADM, we clarify the cosmological requirement for the dark photon portal.
In Sec.\,\ref{sec:model}, we construct a composite ADM model in a bottom-up approach.
The final section is devoted to our conclusions.

\section{ADM thermal history and cosmological constraints on dark photon}
\label{sec:constraint}
In this section, we discuss the thermal history of ADM where the dark photon plays a crucial role.
We also derive cosmological constraints on the dark photon parameters.

\subsection{ADM sector}
Before discussing constraints on the dark photon, let us overview the models of composite ADM.
For that purpose, we consider a SU$(N_{c})_{D}$ gauge dynamics, which is referred to as the QCD$^{\prime}$ in the following.
There are $N_{f}$-flavors of vector-like dark quarks $(Q'_{i}, \bar{Q}'_{i})$ ($i = 1 \cdots N_{f}$) 
with $B - L$ charges of $(q_{B - L}, - q_{B - L})$.
Hereafter, fermions are taken to be the two-component Weyl fermion.
We assume that the masses of the dark quarks, 
\begin{eqnarray}
\label{eq:mass}
{\cal L}_{\rm mass} = \sum_{i} m_{Q_{i}} Q_{i} \bar{Q}_{i} + {\rm h.c.}  \,,
\end{eqnarray}
are smaller than the dynamical scale of the QCD$^{\prime}$, $\Lambda_{\rm QCD'}$.

Below the dynamical scale, we assume that the QCD$^{\prime}$ exhibits a confinement where the dark quarks are confined into dark mesons and dark baryons.
By assuming spontaneous chiral symmetry breaking, we expect that the lightest mesons are the pseudo-Nambu-Goldstone modes, i.e., the dark pions.
The dark pions obtain masses of $m_{\pi'} = {\cal O}(\sqrt{m_{Q} \Lambda_{\rm QCD'}})$, 
which are smaller than the dark baryon masses of $m_{b'} = {\cal O}(\Lambda_{\rm QCD}')$ for $m_{Q} \ll \Lambda_{\rm QCD'}$.
As the dark baryons carry $B - L$ charges, the lightest ones are good candidates for ADM.%
\footnote{We assume that the lightest dark baryons are the ones with the lowest spin, 
while the detailed mass spectrum does not change the following discussion qualitatively.}
The annihilation cross section of the dark baryons into the dark mesons is quite large due to the strong dynamics, 
with which the symmetric part of the DM relic is negligibly small~\cite{Graesser:2011wi, Iminniyaz:2011yp, Bell:2014xta, Baldes:2017gzw, Baldes:2017gzu}.
As a result, the DM abundance is naturally dominated by the asymmetric component.

In our scenario, we assume that the $B-L$ symmetry is softly broken by right-handed neutrino masses, $M_{R}$, which carry a $B - L$ charge of $-2$.%
\footnote{One can gauge the $B-L$ symmetry, which is spontaneously broken by a vacuum expectation value of a scalar whose $B - L$ charge is $-2$.}
The right-handed neutrinos couple to the SM particles via
\begin{eqnarray}
{\cal L}_{N \mbox{-} {\rm SM}} = \frac{1}{2} M_{R} \bar{N}_{R} \bar{N}_{R} +  y_{N} H L \bar{N}_{R} + {\rm h.c.} \, ,
\end{eqnarray}
which trigger the seesaw mechanism.
Here, $H$ and $L$ denote the SM Higgs and the lepton doublets, respectively.
The Yukawa coupling is related to the light neutrino masses via $y_N^2 \sim m_\nu M_R/v_{\rm EW}^2$ with $v_{\rm EW}$ being the vacuum expectation value of the Higgs.
The $B - L$ (i.e., baryon) asymmetry is generated by thermal leptogenesis when the cosmic temperature 
is around the right-handed neutrino mass, $T \sim M_{R} \gtrsim 10^{10}$\,GeV~\cite{Giudice:2003jh, Buchmuller:2005eh, Davidson:2008bu}.

Once the $B-L$ asymmetry is generated, part of it is propagated into the dark sector through the portal interaction,
\begin{eqnarray}
\label{eq:portal}
{\cal L}_{B - L \, {\rm portal}} = \frac{1}{M_{*}^{n}}{\cal O}_{D} {\cal O}_{\rm SM} + {\rm h.c.} \,,
\end{eqnarray}
where ${\cal O}_{D}$ (${\cal O}_{\rm SM}$) is a $B - L$ charged but gauge invariant operator consisting of the dark (SM) sector fields.
Here, $M_{*}$ denotes a portal scale with $n + 4$ ($n\in {\mathbb N}$) being the mass dimension of the operator, ${\cal O}_{D} {\cal O}_{\rm SM}$.
We remark that  the portal operator,  ${\cal O}_{D} {\cal O}_{\rm SM}$, may carry a net $B - L$ charge of $- 2 m$ ($m\in {\mathbb Z}$).
In such a case, the portal scale has a dependence on $M_R$ as $M_{*}^{n} \sim M_{R}^{m} M_{*}^{\prime n - m}$ where $M_{*}'$ 
is a $B-L$ neutral mass parameter.
The ADM scenarios with $B - L$ neutral portal operators (i.e., $m = 0$) were discussed in the literature (e.g., Refs.\,\cite{Kaplan:2009ag, Fukuda:2014xqa}).

The portal interaction eventually decouples at the cosmic temperature of
\begin{eqnarray}
T_{D} \sim M_{*} \left( \frac{M_{*}}{M_{\rm PL}} \right)^{1 / (2n-1)} \,,
\end{eqnarray}
where $M_{\rm PL} \simeq 2.4 \times 10^{18}$\,GeV denotes the reduced Planck scale.
For a successful ADM scenario, $T_D$ is required to be lower than $M_R$.
In the following, we also assume that $T_D$ is higher than the electroweak scale, i.e., the portal interaction decouples before the Sphaleron process decouples.
After the portal interaction decouples, the $B - L$ number is conserved independently in the SM sector and in the dark sector. 
The DM particle is quasi-stable as the decay rate through the portal interaction is suppressed by powers of $\Lambda_{\rm QCD'} / M_{*}$.
In the ADM models with a large annihilation cross section, the DM mass is determined by the ratio of  the $B - L$ asymmetries between the DM and the SM sectors, 
$A_{\rm DM} / A_{\rm SM}$, that is,
\begin{eqnarray}
\label{eq:DMmass}
m_{\rm DM} \simeq 5 \, {\rm GeV} \times \frac{30 A_{\rm SM}}{97 A_{\rm DM}} \,.
\end{eqnarray}
Here, we used the ratio between $A_{\rm SM}$ and the baryon asymmetry observed today, $A_{\rm SM}/A_B = 97/30$~\cite{Harvey:1990qw}.
The value of $A_{\rm SM}/A_{\rm DM}$, which is typically of ${\cal O}(1)$ [as we will see later, e.g., in Eq.\,\eqref{eq:ADMASM}], depends on the dark sector model.
The DM mass in the GeV range can be naturally explained by the dynamical scale of $\Lambda_{\rm QCD'} = {\cal O}(1)$\,GeV.
This is another advantage of the composite ADM models.

\subsection{Dark photon portal}
In the composite ADM models, we assume a strong gauge dynamics in the dark sector.
Thus, the dark sector entropy is sizable in the early Universe, since the dark sector is thermally connected to the SM sector via the portal interaction.
Thus, if the dark pions are stable, they overclose the Universe or contribute to the effective number of neutrino degrees of freedom, $N_{\rm eff}$, too much, depending on their masses~\cite{Blennow:2012de}.
To evade these problems, we introduce a dark photon of the U$(1)_{D}$ gauge interaction, referred to as the QED$^{\prime}$, 
under which the dark quarks are charged.
In the presence of the dark photon, the dark pions can annihilate or decay into the dark photons, which makes 
the dark pion harmless.

The mere introduction of the massless dark photon does not solve the problem, since 
it also contributes to $N_{\rm eff}$ too much.
To avoid this problem, we further assume that the dark photon has kinetic mixing with the SM photon and becomes massive by a Higgs mechanism in the dark sector:
\begin{eqnarray}
\label{eq:kinetic}
{\cal L}_{A'\mbox{-}A} =  \frac{\epsilon}{2} F_{\mu\nu} F'^{\mu\nu} + \frac{1}{2} m_{\gamma'}^{2} A'_{\mu} A'^{\mu} \,.
\end{eqnarray}
Here, $F$ and $F'$ are the field strengths of the SM photon $A$ and the dark photon $A'$, respectively, 
and $m_{\gamma'}$ denotes the mass of the dark photon.  
Through the kinetic mixing parameterized by $\epsilon$, the massive dark photon decays into SM fermions with a decay rate,
\begin{eqnarray}
\label{eq:decay}
\Gamma_{\gamma'} = N_{\rm ch} \frac{1}{3}{\epsilon^{2} \alpha} m_{\gamma'} \simeq 0.3 \, {\rm s}^{-1} \times N_{\rm ch} \left( \frac{\epsilon}{10^{-10}} \right)^{2} \left( \frac{m_{\gamma'}}{100 \, {\rm MeV}} \right) \,.
\end{eqnarray}
Here, $\alpha$ denotes the QED fine-structure constant.
When the dark photon mass is lighter than twice of the muon mass, it decays only into a pair of the electron and the positron, and hence, $N_{\rm ch} = 1$.

Now, let us summarize the thermal history.
Above the decoupling temperature of the portal interaction, $T_D$, the dark sector and the SM sector are in thermal equilibrium and the 
$B-L$ asymmetry is distributed in the two sectors.

Below $T_D$, two sectors evolve independently.
In the dark sector, the confinement of the strong gauge dynamics takes place at the temperature of $T_{\rm QCD'} \sim \Lambda_{\rm QCD'}$.
DM (i.e., the lightest dark baryon) annihilates into the dark mesons with a very large cross section of ${\cal O}(4\pi/m_{\rm DM}^2)$,
with which the symmetric component of DM is erased and only the asymmetric component is left over.
The U$(1)_D$ charged dark pions also annihilate into the dark photons with a cross section of ${\cal O}(4\pi \alpha'^2/m_{\pi'}^2)$
with $\alpha'$ being the QED' fine-structure constant.
This cross section is large enough to make the relic dark pions a subdominant component of DM for $\alpha' \sim \alpha$.%
\footnote{The relic density of the dark pions is subdominant for $m_{\pi'}/\alpha' < {\cal O}(100)$\,TeV.}
Note that the relic density of the neutral pions is also suppressed when the neutral pions are in chemical equilibrium with the charged pions through the inelastic scattering.%
\footnote{If the corresponding chiral symmetry is anomalous for the QED' as in the case of the SM,
the lightest neutral pions decay into the dark photons with a short lifetime.}
The dark photon eventually decays into the SM fermions via the kinetic mixing, so that the initial entropy of the dark sector is transferred to the SM sector.

To realize the above thermal history, we arrange the masses so that
\begin{eqnarray}
\label{eq:massorder}
2 \times m_{e}  < m_{\gamma'} < m_{\pi'} < m_{\rm DM} \,,
\end{eqnarray}
where $m_{e}$ denotes the electron mass.
The first inequality is required to allow the decay of the dark photon.
The second and the third inequalities are required for the annihilations of the charged dark pions and DM. 
As the DM mass is of ${\cal O}(1)\,$GeV, we assume that $m_{\pi'}$ and $m_{\gamma'}$ are in the sub-GeV range.

In the following numerical analysis, we take the $N_{f} = 2$ and $N_{c} = 3$ case, which is the minimal choice as studied in Sec.\,\ref{sec:case2}.
It should be noted that the derived constraints in the following analysis is not significantly changed for a composite ADM model with a different gauge group and/or a different number of the flavors.
Furthermore, our analyses also apply to a more generic dark sector (not necessary the composite ADM) models that have a dark photon portal, with straightforward modifications.

\subsection{Dark photon recoupling}
The most stringent constraints on the dark photon property comes from the constraints on $N_{\rm eff} = 3.15 \pm 0.23$~\cite{Ade:2015xua}
by the precise measurements of the cosmic microwave background (CMB).
Through the decay and the inverse decay of the dark photon, the dark sector recouples to the SM sector at the low cosmic temperature.
We define the recoupling scale factor, $a_{\rm th}$, by
\begin{align}
\label{eq:ath}
3 H (a_{\rm th}) = \frac{K_{1} (m_{\gamma'} / T_{\gamma'})}{K_{2} (m_{\gamma'} / T_{\gamma'})} \Gamma_{\gamma'} \,,
\end{align}
where $H$ denotes the Hubble expansion rate, $K_{n}$ denotes the $n$th order modified Bessel function of the second kind, and the dark photon decay rate at rest, $\Gamma_{\gamma'}$, is given by Eq.~\eqref{eq:decay}.

We approximate the evolution of the dark photon temperature $T_{\gamma'}$ as a function of the scale factor, $a$, by,
\begin{align}
\label{eq:Tgam}
T_{\gamma'}
=
\begin{cases}
		\displaystyle{\frac{a_{\rm QCD'}}{a}} T_{\rm QCD'} & \text{for } a < a_{\rm QCD'}  \,, \\
		T_{\rm QCD'} & \text{for } a_{\rm QCD'} \leq a < a_{\rm F} \,, \\
		\displaystyle{\frac{a_{\rm F}}{a}} T_{\rm QCD'} & \text{for } a_{\rm F} \leq a \,.
	\end{cases}
\end{align}
Here $a_{\rm QCD'}$ denotes the scale factor at the confinement. 
During the period of the dark hadron decoupling, the dark photon temperature does not 
scale by $a^{-1}$ as the dark hadron energy density is transferred to the dark photon.
Since the details of the QCD$^{\prime}$ confinement are not tractable, we simply assume that the dark photon temperature does not change 
during the dark hadron decoupling.
Long after the dark hadron decoupling, on the other hand, the dark photon temperature 
again scales by $a^{-1}$.
The normalization of the dark photon temperature well below $T_{\rm QCD'}$ can be reliably estimated by 
using the entropy conservation before and after the QCD$^\prime$ confinement,
which leads to $a_{\rm F} = (41 / 3)^{1/3} a_{\rm QCD'}$.
Here, we count all the degrees of freedom including the QED$^{\prime}$ breaking Higgs before the QCD$^{\prime}$ confinement,
while we count only the dark photon after the confinement.
In the following numerical analysis, we take $T_{\rm QCD'} = 10 \times T_{\rm QCD}$ with the SM QCD transition temperature $T_{\rm QCD} = 170$\,MeV, 
although the result does not change as long as $T_{\rm QCD'}/T_{\rm QCD} = {\cal O}(10\mbox{--}100)$.

We also estimate the cosmic temperature of the SM sector, $T$, as a function of the scale factor by the entropy conservation, following Ref.\,\cite{Saikawa:2018rcs}.
Here note that $T = T_{\gamma'}$ for $a < a_{D}$, where $a_{D}$ denotes the scale factor when the portal operator decouples.
The impact of the resultant dark photon density on $N_{\rm eff}$ depends on whether the reheating temperature 
of the SM sector by the dark photon recoupling is above or below the neutrino decoupling temperature, $T_{\nu \mbox{-} {\rm dec}} \simeq 3$\,MeV.
This is because the dark photon energy primarily heats up the electromagnetic particles only.
To see the temperature at the recoupling, we define $T_{\rm cr}$ as 
\begin{align}
\label{eq:eq4}
	\rho _{\rm SM} (a_{\rm th}) + \rho_{\gamma'} (a_{\rm th}) = \rho _{{\rm SM} + \gamma'} (T_{\rm cr}) \,,
\end{align}
where the left-hand (right-hand) side denotes the energy densities before (after) the recoupling.
Here, the energy density of the dark photon before the recoupling, $\rho_{\gamma'} (a_{\rm th})$, is evaluated with the following distribution function:%
\footnote{This is valid when double Compton scattering and bremsstrahlung of the dark proton become inefficient before the dark photon becomes non-relativistic.
If the dark photon is in thermal bath when it becomes non-relativistic, the entropy conservation requires $\rho_{\gamma'} \propto a^{3} / \ln(a)$.
A similar situation can be found in the freeze-out of self-interacting DM through a $3 \to 2$ process~\cite{Carlson:1992fn}.
The resultant $\rho_{\gamma'} (a_{\rm th})$ is different from our evaluation only by a small logarithmic factor.}
\begin{align}
f_{\gamma'}(p, a, T_{\mathrm{F}}) = \frac{1}{\exp( \sqrt{m_{\gamma'}^{2} + (a / a_{\mathrm{F}})^{2} p^{2}}/T_{\mathrm{F}}) - 1} \,.
\end{align}
The energy density after the recoupling, $\rho_{{\rm SM}+\gamma'}$, is simply given by the one in thermal equilibrium with a common temperature $T_{\rm cr}$.
For $T_{\rm cr} > T_{\nu \mbox{-} {\rm dec}}$, we judge that the dark photon recoupling reheats the whole SM sector.
Otherwise, we judge that the recoupling reheats only the electron and the photon.
In the latter case, we estimate the temperature of the electron$+$photon system, $T_{\rm com}$, by
\begin{align}
\label{eq:eq5}
\rho _{\gamma + e} (a_{\rm th}) + \rho_{\gamma'} (a_{\rm th}) = \rho _{\gamma + e + \gamma'} (T_{\rm com}) \,,
\end{align}
which differs from $T_{\rm cr}$,
while we assume that the neutrino temperature is not affected by the decay of the dark photon.

\subsubsection{Dark photon recoupling above the neutrino decoupling temperature: $T_{\rm cr} > T_{\nu \mbox{-} dec}$}
\label{sec:r1}
\begin{figure}[t]
\begin{center}
  \includegraphics[width=.55\linewidth]{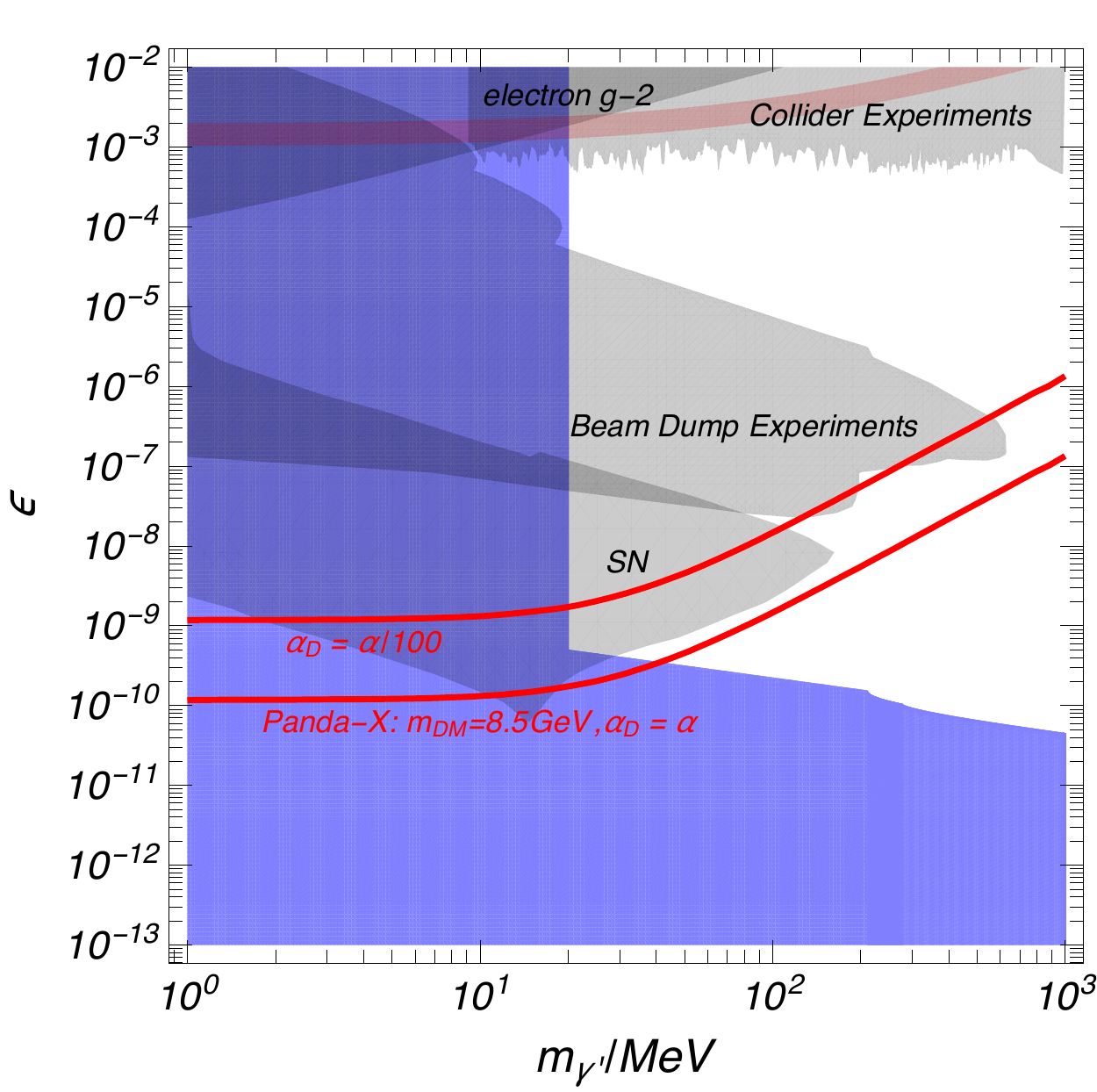}
 \end{center}
\caption{\sl \small
Constraints on the dark photon parameters.
The blue shaded regions are excluded by the cosmological constraints discussed in Secs.\,\ref{sec:r1} (left) and \ref{sec:r2} (bottom).
We take $T_{\rm QCD'} = 1$\,GeV and $a_{\rm F} / a_{\rm QCD'} = 3$, although the result barely depends on their values.
The gray shaded regions are excluded by SN 1987A~\cite{Chang:2016ntp, Chang:2018rso}, beam dump experiments, and collider experiments~\cite{Bauer:2018onh}.
The red lines show the upper limit on $\epsilon$ at $90$\%\,CL for $m_{\rm DM} = 8.5$\,GeV from the DM direct detection experiment (see Sec.~\ref{sec:case2}).
We set $\alpha_{D} = \alpha$ (lower) and $\alpha_{D} = \alpha/100$ (upper).
In our analysis, we use the Maxwell velocity distribution with the velocity dispersion of $v_{0} = 220$\,km/s, which is truncated at the Galactic escape velocity, $v_{\rm esc} = 544$\,km/s.
The local circular velocity is also fixed to be $v_{\rm circ} = 220$\,km/s with the peculiar motions of the Earth being neglected.
We use a conventional value of the local DM density, $\rho_{\rm DM} = 0.3\,$GeV/cm$^{3}$, assuming that half of the total DM consists of $p'$.
We note that the direct detection constraint is sensitive to the DM mass, while the other constraints are not.
}
\label{fig:constraint}
\end{figure}

When the reheated temperature of the SM sector by the dark photon recoupling exceeds the neutrino decoupling temperature, i.e., $T_{\rm cr} > T_{\nu \mbox{-} {\rm dec}}$,
most energy of the dark-photon is re-distributed among the photons, the electrons, and the neutrinos.
This situation corresponds to the kinetic mixing parameter of
\begin{eqnarray}
\label{eq:KM}
\epsilon \gtrsim 10^{-10} \times 
\left(
\frac{10\,{\rm MeV}}{m_{\gamma'}}
\right)^{1/2}\ .
\end{eqnarray} 
Even in this case, a portion of the thermalized dark photons releases its energy at the temperature below $T_{\nu \mbox{-} {\rm dec}}$, which reheats only the 
electrons and the photons.
Thus, such late-time energy injection affects the neutrino-to-photon temperature ratio, $T_{\nu} / T_{\gamma}$.
By considering the entropy conservation in the electron+photon+dark photon plasma and that in the neutrino plasma independently, one finds
\begin{align}
	\label{eq:eq1}
		\frac{T_{\nu}}{T_{\gamma}}= \left(\frac{4}{11}\right)^{1/3} \left(1+\frac{45}{11 \pi ^2}\frac{s_{\gamma '}(T_{\nu\mbox{-}\rm dec})}{T_{\nu\mbox{-}\rm dec}^3}\right)^{-1/3} \,.
\end{align} 
Here, $s_{\gamma'} (T_{\nu\mbox{-}\rm dec})$ is the entropy density of the dark photon at $T_{\nu\mbox{-}\rm dec}$.
As a result, the effective number of neutrino types is changed to  
\begin{align}
\label{eq:eq2}
	N_{\rm eff} = \left( 1 + \frac{45}{11 \pi^{2}}\frac{s_{\gamma'}(T_{\nu\mbox{-}\rm dec})}{T_{\nu\mbox{-}\rm dec}^3}\right)^{-4/3} N_{\nu}^{(\rm SM)} \,,
\end{align}
where $N_{\nu}^{(\rm SM)} = 3.046$~\cite{Mangano:2005cc} (3.045 in the recent analysis~\cite{deSalas:2016ztq}) is the SM value. 
From the CMB observation, we find that the dark photon mass is bounded from below,
\begin{align}
\label{eq:eq3}
	m_{\gamma'} \gtrsim 20\, \mathrm{MeV} \,.
\end{align}
We note that the constraint does not depend on $\epsilon$ in this case 
since the criterion depends only on the entropy density of the dark photon at the neutrino decoupling temperature 
(see Fig.\,\ref{fig:constraint}).

\subsubsection{Dark photon recoupling below the neutrino decoupling temperature: $T_{\rm cr} < T_{\nu \mbox{-}\rm dec}$}
\label{sec:r2}
When the kinetic mixing parameter is  smaller than Eq.\,\eqref{eq:KM}, the dark photon recoupling reheats only the electron and the photon thermal bath,
while the neutrino temperature does not get contributions from the dark photon.
In this case, the entropy conservation of the neutrino plasma for $a > a_{\nu \mbox{-}\rm dec}$ and that of the electron+photon+dark photon plasma leads to
the neutrino-to-photon temperature ratio after the recoupling:
\begin{align}
	\frac{T_\nu}{T_\gamma} = \left(\frac{4}{11}\right)^{1/3} \frac{T_{\nu\mbox{-}\rm dec}}{T_{\rm com}}
\frac{a_{\nu\mbox{-}\rm dec}}{a_{\rm th}} \,.
\end{align}
Accordingly, the effective number of neutrino degrees of freedom is estimated to be
\begin{align}
N_{\mathrm{eff}} = \left( \frac{T_{\nu\mbox{-}\rm dec}}{T_{\rm com}} \right)^{4} \left( \frac{a_{\nu\mbox{-}\rm dec}}{a_{\rm th}} \right)^{4} N_{\nu}^{(\rm SM)} \,.
\end{align}
The lower blue shaded region of Fig.\,\ref{fig:constraint} shows the resultant constraint.
Roughly speaking, the upper limit on $\epsilon$ for a given $m_{\gamma'}$ corresponds to the dark photon lifetime of ${\cal O}(1)$\,s,
and hence, to the dark photon decay at the neutrino decoupling temperature.
In the figure, we take account of the muon, the charged pion, and the charged Kaon decay channels 
 in $\Gamma_{\gamma'}$ in addition to the electron one, when those modes are kinematically allowed.

\section{Bottom-up construction of a composite $B - L$ ADM model}
\label{sec:model}
In this section, we discuss a minimal model of a composite $B - L$ ADM, which achieves the thermal history discussed in the previous section.
For $N_f = 1$, an operator ${\cal O}_{D}$ charged under the $B - L$ symmetry is also charged under U$(1)_D$,
and hence, no $B-L$ portal interaction is allowed.
Thus we take $N_{f} = 2$ as a minimal model.

\subsection{$N_{c} = 2$ case}
\label{sec:case1}

\begin{table}[t]
\begin{center}
\begin{tabular}{|Y|c|c|c|c|}
\hline
& SU$(2)_{D}$ & $B - L$ & U$(1)_{D}$ \\
\hline
\rule{0pt}{2.3ex} $Q_{1}$ & ${\bf 2}$ & $q_{B - L}$ & 1/2 \\
\hline
\rule{0pt}{2.3ex} $\bar{Q}_{1}$ &${\bf 2}$ & $- q_{B - L}$ & -1/2\\
\hline
\rule{0pt}{2.3ex} $Q_{2}$ &${\bf 2}$ &$q_{B - L}$ & -1/2\\
\hline
\rule{0pt}{2.3ex} $\bar{Q}_{2}$ &${\bf 2}$&$- q_{B - L}$ & 1/2\\
\hline
\end{tabular}
\end{center}
\caption{\sl \small
The charge assignment of the minimal model with $N_{c} = 2$ and $N_{f} = 2$.
The QED$^{\prime}$ charges of dark quarks are normalized to be $\pm 1/2$ without loss of generality.
}
\label{tab:Nc2}
\end{table}%
In this section, we discuss a model with $N_{c} = 2$ and $N_{f} = 2$.
In Table\,\ref{tab:Nc2}, we show the charge assignment of the dark quarks.
In this case, the dark pions are 
\begin{eqnarray}
\pi^{\prime 0} \propto Q_{1} \bar{Q}_{1} - Q_{2} \bar{Q}_{2}  \,,
\quad
\pi^{\prime +}  \propto Q_{1} \bar{Q}_{2} \,,
\quad
\pi^{\prime -}  \propto {Q}_{2} \bar{Q}_{1}  \,.
\end{eqnarray}
and the lightest baryons%
\footnote{In this case the lightest baryons are also Nambu-Goldstone modes due to the enhanced symmetry breaking patter in the chiral limit, U$(4) [\supset$ SU$(2) \times$SU$(2) \times$U$(1)] \to$ USp$(4) [\supset$ SU$(2) \times$U$(1)$].
Even if the masses of the lightest baryons are degenerate with those of the dark pions, the thermal history  discussed above does not change as long as the dark baryon annihilation into the dark pions is efficient.
}
 are
\begin{eqnarray}
b \propto Q_{1} Q_{2} \,, \quad  \bar{b} \propto \bar{Q}_{1} \bar{Q}_{2} \,.
\end{eqnarray}

The lowest dimensional portal interaction is
\begin{eqnarray}
{\cal L}_{N} = \frac{1}{M_{*}^{\prime 2}} \bar{Q}_{1} \bar{Q}_{2} \bar{N}_{R}^{2}  + {\rm h.c.} \,,
\end{eqnarray}
which requires $q_{B - L} = 1$.%
\footnote{Terms like $Q_{1}^{\dagger} Q_{2}^{\dagger} \bar{N}_{R}^{2}
/ M_{*}^{\prime 2}$ are also possible.
Hereafter we omit such parity conjugated terms for the sake of notational simplicity, as they do not alter the discussion.}
By integrating out $\bar{N}_{R}$, we obtain 
\begin{eqnarray}
\label{eq:portal21}
{\cal L}_{B - L \, {\rm portal}} = \frac{y_{N}^{2}}{M_{R}^{2} M_{*}^{\prime 2}} (\bar{Q}_{1} \bar{Q}_{2}) (L H)^{2} 
+ {\rm h.c.}\ , 
\end{eqnarray}
and hence, $M_{*}^{4} = M_{R}^{2} M_{*}^{' 2}/ y_{N}^{2}$ in eq. (2.3).
A drawback of this charge assignment is that it allows
\begin{eqnarray}
\label{eq:washout}
{\cal L}_{B - L \, {\rm mass}} = M_{R} (Q_{1} Q_{2})  + M_{R}^{\dagger} (\bar{Q}_1 \bar{Q}_2 ) +  {\rm h.c.} \,,
\end{eqnarray}
which result in the dark quarks masses of ${\cal O}(M_{R})$.
Thus, the minimal charge assignment contradicts with the assumption of the composite model.

We may take $q_{B - L} = \pm 1/2$ to avoid the unwanted mass term in Eq.\,\eqref{eq:washout}.
In this case, the lowest dimensional portal interaction is
\begin{eqnarray}
\label{eq:portal22}
{\cal L}_{B - L \, {\rm portal}} = \frac{y_{N}^{2}}{M_{R}^{2} M_{*}^{\prime 5}} (\bar{Q}_{1} \bar{Q}_{2})^{2} (L H)^{2} 
+ {\rm h.c.}
\end{eqnarray}
This portal operator is a valid choice for the purpose of the ADM scenario.
As we will see shortly, however, we have a model with 
a simpler portal interaction for $N_c = 3$. 
Hence, we do not pursue this possibility further,
although it is phenomenologically consistent.%
\footnote{From the point of view of the ultraviolet completion, the portal interaction in Eq.\,\eqref{eq:portal22}
requires more complicated structure than the one required for the portal interaction in Eq.\,\eqref{eq:portalN}.
}
We also stress that the constraints on the dark photon parameter space in the previous section is not significantly changed 
for the model with $N_c =2$ and $N_f =2$, although they are derived for $N_c = 3$ and $N_f= 2$.

\subsection{$N_{c} = 3$ case}
\label{sec:case2}

\begin{table}[t]
\begin{center}
\begin{tabular}{|Y|c|c|c|c|}
\hline
&SU$(3)_{D}$ & $B - L$& U$(1)_{D}$ \\
\hline
\rule{0pt}{2.3ex} $Q_{1}$ & ${\bf 3}$ & $q_{B - L}$ & 2/3 \\
\hline
\rule{0pt}{2.3ex} $\bar{Q}_{1}$ & ${\bf \bar{3}}$ & $-q_{B - L}$ & -2/3 \\
\hline
\rule{0pt}{2.3ex} $Q_{2}$ & ${\bf 3}$ & $q_{B - L}$ & -1/3\\
\hline
\rule{0pt}{2.3ex} $\bar{Q}_{2}$ & ${\bf \bar{3}}$ & $-q_{B - L}$ & 1/3 \\
\hline
\end{tabular}
\end{center}
\caption{\sl \small
The charge assignment of the minimal model for $N_{c} = 3$ and $N_{f} = 2$.
The QED$^{\prime}$ charges are assigned so that one of the dark baryon becomes neutral.
}
\label{tab:Nc3}
\end{table}%
Now, let us consider the case with $N_{c} = 3$ and $N_{f} = 2$.
In Table\,\ref{tab:Nc3}, we show the charge assignment of the dark quarks.
As the charge assignment is parallel to the QCD charge, it is apparently free from quantum anomalies.%
\footnote{This model has a similarity to models based on the idea of the mirror matter~\cite{Foot:2003jt, An:2009vq, Farina:2015uea, Lonsdale:2018xwd}.
In such scenarios, mirror baryons are DM candidates, although the mirror photon is massless.
}
In this case, the dark pions are
\begin{eqnarray}
\pi^{\prime 0} \propto Q_{1} \bar{Q}_{1}  - Q_{2} \bar{Q}_{2} \,,
\quad
\pi^{\prime +}  \propto Q_{1} \bar{Q}_{2} \,,
\quad
\pi^{\prime -}  \propto {Q}_{2} \bar{Q}_{1} \,,
\end{eqnarray}
and the dark baryons are
\begin{eqnarray}
p' \propto Q_{1} Q_{1} Q_{2} \,, \quad 
\bar{p}' \propto \bar{Q}_{1} \bar{Q}_{1} \bar{Q}_{2} \,, \quad 
n' \propto Q_{1} Q_{2} Q_{2} \,, \quad 
\bar{n}' \propto \bar{Q}_{1} \bar Q_{2} \bar Q_{2} \,.
\end{eqnarray}
We summarize hadron mass formulas in the appendix\,\ref{sec:hadronmass}.
We emphasize that the QED$^{\prime}$ charge assignment in Table\,\ref{tab:Nc3} is 
the unique choice (up to trivial normalization) that makes one of the dark baryon neutral and allows the following portal interaction.

The lowest dimensional portal interaction is
\begin{eqnarray}
\label{eq:portalN}
{\cal L}_{N \mbox{-} D} =  \frac{1}{M_{*}^{\prime 2}} (\bar{Q}_{1} \bar{Q}_{2} \bar{Q}_{2}) \bar{N}_{R} + {\rm h.c.} \,,
\end{eqnarray}
which requires $q_{B - L} = 1/3$.
Below the mass scale of $M_{R}$, the above portal interaction results in
\begin{eqnarray}
\label{eq:portal3}
{\cal L}_{B - L \, {\rm portal}} =  
 \frac{y_{N}}{M_{*}^{\prime 2} M_{R}} (\bar{Q}_{1} \bar{Q}_{2} \bar{Q}_{2} ) L H + {\rm h.c.} \ ,
\end{eqnarray}
and hence, $M_{*}$ in Eq.\,\eqref{eq:portal} should be identified as $(M_{*}^{\prime 2} M_{R} / y_{N})^{1/3}$.

As we mentioned earlier, we assume that $T_{D}$ is below the right-handed neutrino mass scale%
\footnote{It follows that $M_{*}^{\prime} \lesssim 10 M_{R} \times (m_{\nu} / 0.1 \, {\rm eV})^{1/4}$.
}
 and is above the decoupling temperature of the Sphaleron process.
In this case, the ratio of the $B - L$ asymmetries between the dark and the SM sectors is~\cite{Fukuda:2014xqa}
\begin{eqnarray}
\label{eq:ADMASM}
\frac{A_{\rm DM}}{A_{\rm SM}} = \frac{44}{237} \,.
\end{eqnarray}
As a result, we find that the mass of DM is $m_{\rm DM} = 8.5$\,GeV [see Eq.~\eqref{eq:DMmass}], 
for which we take $\Lambda_{\rm QCD'} \sim 10 \times \Lambda_{\rm QCD}$ with $\Lambda_{\rm QCD} \sim 200$\,MeV denoting the QCD dynamical scale.
We consider a dark pion mass of ${\cal O}(10 \text{--} 100)$\,MeV or larger, which corresponds to a quark mass of ${\cal O}(1)$\,MeV or larger.
The dark neutron can be heavier or lighter than the dark proton depending on the quark mass parameters $m_{1}$ and $m_2$ [see Eq.\,\eqref{eq:massdiff}].

An interesting feature of the portal interaction [see Eq.\,\eqref{eq:portal3}] is that it leads to a decay of the dark nucleon into a pair of the dark pion and the SM neutrino.
Although the predicted lifetime of the dark nucleon is much longer than the age of the Universe,
the dark nucleon decay is constrained by the measurements of the neutrino flux by the Super-Kamiokande (SK) experiment,
which puts the lower limit on the portal scale from below as $M_{*} \gtrsim 10^{8.5}\,$GeV~\cite{Fukuda:2014xqa} (see also Ref.\,\cite{Covi:2009xn}).

The dark proton does not mix with the dark neutron when the dark Higgs boson has a U$(1)_{D}$ charge of $-2$, since the ${\mathbb Z}_{2}$ subgroup of U$(1)_{D}$ remains unbroken. 
When the dark Higgs boson charge is $-1$, on the other hand, the dark proton slightly mixes with the dark neutron. In the following, we consider the model with the dark Higgs charge of $-2$,
although the case with the dark Higgs charge of $-1$ is also a viable option as discussed in appendix B.

As another interesting feature of the model, the dark proton has a coupling to the SM fermions through the dark photon,
with which DM can be searched for.
As the dark neutron does not couple to the dark photon, the expected event rate of the DM direct detection depends on the dark proton fraction in DM.
The dark proton fraction is determined by the dark nucleon inelastic scattering with the dark pion, which freezes out when the dark pions annihilate into the dark photons, $T_{\gamma'} \sim m_{\pi'} / 20 \text{--} 30$.
Resultantly, the dark proton fraction is given by 
\begin{eqnarray}
\label{eq:p'ratio}
\frac{n_{p'}}{n_{n'} + n_{p'}} = \frac{1}{\exp[- (m_{n'} - m_{p'}) / T_{\gamma'}] + 1} \,.
\end{eqnarray}
Since the $n'$--$p'$ mass difference, $m_{n'} - m_{p'} = {\cal O} (m_{1, 2})$, is basically smaller than the dark pion mass, $m_{\pi'} = {\cal O}(\sqrt{m_{1, \, 2} \Lambda_{\rm QCD'}})$ (see appendix\,\ref{sec:hadronmass}), we consider that $p'$ accounts for half of DM.
\footnote{If the dark photon mass is smaller than the dark deuterium binding energy, the dark nucleosynthesis could proceed~\cite{Krnjaic:2014xza, Detmold:2014qqa, Detmold:2014kba, Wise:2014ola, Hardy:2014mqa, Gresham:2017cvl, Gresham:2017zqi} and significantly change direct detection signals.
If one estimates the dark deuterium binding energy as $B_{d'} \sim B_{d} \times \Lambda_{\rm QCD'} / \Lambda_{\rm QCD}$ with the SM value, $B_{d} \simeq 2.2$\,MeV, the direct detection constraint would differ from our analysis for $m_{\gamma'} \lesssim 20 \, {\rm MeV}$.}

Following the analysis in Ref.\,\cite{DelNobile:2015uua}, we place the upper bound on $\epsilon$ on the dark photon parameter from the $54$ ton$\times$day exposure of PandaX-II~\cite{Cui:2017nnn} in Fig.\,\ref{fig:constraint} (red lines).
With this exposure, no signal candidates were observed while the expected background in the signal region was $1.8 \pm 0.5$.
This leads to an upper limit of 0.63 signal events in the signal region at $90$\%\,CL.
Similar constraints are expected from the results of LUX~\cite{Akerib:2016vxi} and XENON1T~\cite{Aprile:2017iyp}.
The direct detection experiment constraint is severer than that from SN 1987A for the QED$^{\prime}$ fine-structure constant $\alpha_{D} = \alpha$.
For $\alpha_D = \alpha$ and $m_{\gamma'} \lesssim 100$\,MeV, large portion of the parameter region can be tested by
future experiments such as XENONnT~\cite{Aprile:2015uzo}, LZ~\cite{Mount:2017qzi}, and Darwin~\cite{Baudis:2012bc}.
With a light mediator, $m_{\gamma'} \lesssim 100$\,MeV, the nuclear recoil energy spectrum of DM scattering is distinguishable from that of the neutrino background~\cite{Dent:2016wor}.

Finally, let us remark that there can be an additional $B- L$ neutral portal operator,
\begin{eqnarray}
\label{eq:portal4}
{\cal L}_{B - L \, {\rm portal}} = \frac{1}{M_{*}^{3}} (Q_{1} Q_{1} Q_{2}) (L H) + {\rm h.c.}
\end{eqnarray}
When both the portals in Eqs.\,\eqref{eq:portal3} and \eqref{eq:portal4} are effective in the thermal bath at the temperature below $M_{R}$, 
the $B - L$ asymmetry generated by thermal leptogenesis is washed out.
To avoid this problem, we assume that $M_{*}$ in Eq.\,\eqref{eq:portal4} is much larger than that in Eq.\,\eqref{eq:portal3}.

This can be realized, for instance, as follows.
One introduces a scalar quark $\phi$ that transforms as a fundamental representation of SU$(3)_D$ with the QED$^{\prime}$ charge of $2/3$ and $B - L$ charge of $-2/3$.
With this charge assignment, the portal interaction in Eq.\,\eqref{eq:portalN} is generated via
\begin{eqnarray}
{\cal L}_{\phi \mbox{-} D} = y_{\phi N} \, \phi \bar{N}_{R} \bar{Q}_{1} +  y_{\phi Q} \, \phi^{*} \bar{Q}_{2} \bar{Q}_{2} + {\rm h.c.} \,,
\end{eqnarray}
below the scale of the scalar quark mass $M_{\phi} > M_{N}$.
The suppression scale is $M_{*} = {\cal O}(M_{\phi}^{2} M_{N} / y_{N})^{1/3}$.
Meanwhile the portal interaction in Eq.\,\eqref{eq:portal4} is generated via
\begin{eqnarray}
{\cal L}_{\phi \mbox{-} D'} = \frac{1}{M_{*}''} \phi^{*} Q_{1} L H + y_{\phi Q}'  \, \phi \, {Q}_{2} {Q}_{2} \,,
\end{eqnarray}
which leads to a larger suppression scale, $M_{*} = {\cal O}(M_{\phi}^{2} M_{*}'')^{1/3}$, when $M_{*}'' \gg M_{\phi} > M_{N}$.

\section{Conclusions}
Motivated by the composite ADM scenario, we have investigated the viable parameter space of  the dark photon portal, 
with which the entropy of the dark sector is transferred to the SM sector.
As we have seen, the stringent bound comes from the observational constraint on $N_{\rm eff}$, where the bound depends on whether the reheating temperature of the SM sector by the dark photon recoupling is above or below the neutrino decoupling temperature.
If the neutrinos are in  the thermal bath at the dark photon recoupling, the recoupling itself does not affect $N_{\rm eff}$.
Still, the thermalized dark photons affect $N_{\rm eff}$ by heating only electron and photon plasma after the neutrino decoupling.
The observational constraint on $N_{\rm eff}$ places an upper bound on the dark photon mass in this case.
f the neutrinos already decoupled from the thermal bath at the dark photon recoupling, the recoupling directly heats the electrons and the photons and thus changes $N_{\rm eff}$.
We have obtained a lower bound on the kinetic mixing parameter for a given dark photon mass.

In addition, we have constructed a minimal model of composite ADM, which is compatible with the seesaw mechanism and thermal leptogenesis.
It has a QCD-like SU$(3)$ gauge theory and a QED-like U$(1)$ gauge interaction.
As the dark proton is charged under U$(1)_{D}$, our ADM can be tested by direct detection experiments.
We have found that the current direct detection constraint is severer than that from SN 1987A.
A large portion of the parameter space can be tested by future experiments such as XENONnT, LZ, and Darwin.

Our model of ADM has interesting astrophysical implications.
As the QCD$^{\prime}$ is similar to the SM QCD, our ADM would have a cross section similar to that of the SM nucleon, which is ${\cal O}(1) \, {\rm b}$ and constant at low velocity, while diminishes with increasing velocity above $v/c = {\cal O} (10^{-2})$~\cite{Chadwick:2011xwu}.
Such a velocity-dependent cross section could solve issues of cold dark matter structure formation on galactic scales, while satisfying the constraints from galaxy clusters (see, e.g., Ref.~\cite{Tulin:2017ara} for a review).
The nature of our ADM self-scattering and its implications for structure formation are worth investigating.

\begin{acknowledgments}
A. K. thanks Takumi Kuwahara for useful discussions.
This work is supported in part by Grants-in-Aid for Scientific Research from the Ministry of Education, Culture, Sports, Science, and Technology (MEXT) KAKENHI, Japan,  No. 15H05889, No. 16H03991, No. 17H02878  (M. I.), by IBS under the project code IBS-R018-D (A. K.) and by the World Premier International Research Center Initiative (WPI), MEXT, Japan.
\end{acknowledgments}

\appendix

\section{Hadron mass spectrum in the QCD$^{\prime}$ + QED$^{\prime}$ model}
\label{sec:hadronmass}

By analogy to the QCD, the masses of the dark pions are estimated as
\begin{eqnarray}
\label{eq:pi0}
m_{\pi^{\prime 0}}^{2} \simeq m_{\pi^0}^{2} \times \frac{\Lambda_{\rm QCD'}}{\Lambda_{\rm QCD}} \frac{m_{1} + m_{2}} {m_{u} + m_{d}} \,, 
\end{eqnarray}
where $m_{u \, (d)}$ is the SM up-type (down-type) quark mass.
The squared mass difference of the dark pions is given by
\begin{eqnarray}
\label{eq:pi02}
m_{\pi^{\prime\pm}}^{2} \simeq m_{\pi^{\prime 0}}^{2} + \alpha_{D} \Lambda_{\rm QCD'}^{2} \,.
\end{eqnarray}

The average dark (SM) nucleon mass $m_{N' \, (N)}$ is given by
\begin{eqnarray}
m_{N'} \simeq m_{N} \times \frac{\Lambda_{\rm QCD'}}{\Lambda_{\rm QCD}} \,,
\end{eqnarray}
while the nucleon mass difference is given by
\begin{eqnarray}
\label{eq:massdiff}
m_{n'} - m_{p'} \simeq \delta m_{n \mbox{-} p}^{\rm QED}  \times 
\frac{\Lambda_{\rm QCD'}}{\Lambda_{\rm QCD}} 
\times \alpha_D
+ \kappa_{N} (m_{1} - m_{2}) \,.
\end{eqnarray}
Here, $\delta m_{n \mbox{-} p}^{\rm QED} = - 0.178^{+0.004}_{-0.064}$\,GeV and $\kappa_{N} = 0.95^{+0.08}_{-0.06}$ parameterize the electromagnetic and the isospin-violating contributions, respectively~\cite{Walker-Loud:2014iea}.

\section{U$(1)_{D}$ Higgs with a charge of $-1$}
\label{sec:m1}
For the U$(1)_{D}$ Higgs charge of $-1$, Yukawa couplings,
\begin{align}
\label{eq:Higgsportal}
{\cal L}_{{\rm U}(1)_{D} \, {\rm mass}} = y H_{D} Q_{1} \bar{Q}_{2} + \bar{y} H_{D}^{\dagger} \bar{Q}_{1} Q_{2} + {\rm h.c.} \,,
\end{align}
induce mixing between the QED$'$ breaking Higgs and the charged pion.
The charged pion also develops a vacuum expectation value.
It induces dark proton mixing with the dark neutron once the U$(1)_{D}$ symmetry is spontaneously broken.
In this case, the heavier nucleon can decay into the lighter one and the dark photon or the charged leptons, depending on the mass difference $m_{n'} - m_{p'}$ [see Eq.\,\eqref{eq:massdiff}].
If the dark photon channel is kinematically forbidden, the lifetime of the heavier nucleon is of ${\cal O}(10^{10})$\,s for $\epsilon = {\cal O}(10^{-10})$ and its fraction in the whole DM is severely constrained to be smaller than ${\cal O}(10^{-4})$ by the light element abundance~\cite{Poulin:2016anj}.
The fraction is determined by the dark nucleon inelastic scattering with the dark photon, which is induced by the vacuum expectation value of the charge pion.
This interaction decouples when $T_{\gamma'} \sim |m_{n'} - m_{p'}| / 20 \text{--} 30$ if the dark photon decays below this dark photon temperature.
The resultant fraction is of ${\cal O}(10^{-9})$ [see Eq.\,\eqref{eq:p'ratio}], which evades the above cosmological constraint.
If the dark photon decays before $T_{\gamma'} \sim |m_{n'} - m_{p'}| / 20 \text{--} 30$, the freeze-out temperature is given by $T_{\gamma'} \sim m_{\gamma'} / 20 \text{--} 30$.%
\footnote{If the SM reheating temperature [see Eqs.\,\eqref{eq:eq4} and \eqref{eq:eq5}] is lower than this dark photon temperature, the freeze-out temperature is the dark photon temperature at recoupling.
If double Compton scattering and bremsstrahlung of the dark proton are efficient, the freeze-out of the dark photon becomes non-trivial.
This is because the dark photon temperature drop only slowly as $T_{\gamma'} \propto 1 / \ln a$ after dark photon becomes non-relativistic.}
In this case, the resultant fraction tends to exceed the upper bound from the cosmological constraint.

The charged pion also mixes with the SM Higgs boson through the vacuum expectation value of the charge pion and the Higgs portal coupling, $|H|^{2}|H_{D}|^{2}$.
Resultantly the charged pion decays into the SM fermions.
Such decay modes provide an alternative route (to kinetic mixing) to transfer the entropy in the dark sector to the SM sector (see, e.g., Ref.\,\cite{Kouvaris:2014uoa}).
A detailed discussion will be given elsewhere.

\bibliographystyle{utphys}
\bibliography{draft_arXiv_replacement}

\end{document}